\documentstyle[twoside,fleqn,epsf,espcrc2]{article}
\newcommand{\mrm}{\mathrm}
\newcommand{\figfudge}{}

\title{The Heavy Hybrid Spectrum from NRQCD and the Born-Oppenheimer 
       Approximation\thanks{Talk presented by K.J.~Juge.}}

\author{K.J.~Juge\address{Fermi National Accelerator Laboratory, 
                          P.O.~Box 500, Batavia, Illinois 60510}, 
	J.~Kuti\address{Dept.~of Physics, University of California at San Diego,
                        La Jolla, California 92093-0319}
        and C.J.~Morningstar${}^{\rm b}$}

\begin{document}

\begin{abstract}
The spectrum of heavy-quark hybrids is studied in the leading Born-Oppenheimer (LBO) approximation and using leading-order NRQCD simulations with an improved gluon action on anisotropic lattices.  The masses of four hybrid states are obtained from our simulations for lattice spacings 0.1 fm and 0.2 fm and are compared to the LBO predictions obtained using previously-determined glue-excited static potentials. The consistency of results from the two approaches reveals a compelling physical picture for heavy-quark hybrid states.
\end{abstract}

\maketitle

\section{Introduction}
QCD predicts the existence of hybrid states which contain excited gluon fields. Hybrid mesons with heavy ${\rm\overline{b}b}$ quark pairs are the most amenable to theoretical treatment. They can be studied not only directly by numerical simulation, but also using the Born-Oppenheimer expansion which is our primary guidance for the development of a simple physical picture. The Born-Oppenheimer picture was introduced for the description of heavy hybrid states in Refs.~\cite{hasenfratz} and was applied using hybrid potentials first calculated in lattice QCD in Ref.~\cite{michael}. In this study, we work to leading order in the expansion and neglect higher-order terms involving spin, relativistic, and retardation effects.  We test the accuracy of the Born-Oppenheimer approach by comparison with high-precision results from simulations.  

\section{NRQCD Simulation}

\begin{table}
\caption{
   Simulation parameters for the two lattices.
   \label{table:simparams}}
\begin{center}
\begin{tabular}{llll}\hline
 $(\beta,\xi)$            &  $(3.0,3)$          &  $(2.6,3)$ \\[1mm]
 ${\rm u_s^4}$            &  $0.500$            &  $0.451$   \\
 lattice                  &  $15^3\!\times\!45$ &  $10^3\!\times\!30$ \\
 $\#$ configs, sources    &  $201, 16080$       &  $355, 17040$ \\
 ${\rm r_0/a_s}$          &  $4.130(24)$        &  $2.493(9)$ \\
 ${\rm a_sM_0}$           &  $2.6(1)$           &  $3.9(1)$ \\[1mm]\hline
\end{tabular}
\figfudge
\end{center}
\end{table}

In our NRQCD simulations, we use the improved gauge-field action of Ref.~\cite{peardon} on anisotropic lattices with $a_s/a_t\!=\!3$. The simulation parameters are listed in Table~\ref{table:simparams}. We set the aspect ratio using the tree level anisotropy in all our calculations. The time symmetrized equation with stability parameter $n\!=\!2$ is used for the evolution of the nonrelativistic propagators. The Hamiltonian consists of the leading kinetic energy operator and two other operators to remove ${\mathcal O}(a_t)$ and ${\mathcal O}(a_s^2)$ errors.

Our meson/hybrid operators are constructed as follows. First the spatial links are APE-smeared. Then, these links are used to Jacobi-smear the quark (antiquark) fields and to make the (clover-leaf) chromomagnetic field. The specific operators used to project out each channel are listed in Table~\ref{table:mesonops}. Note that we use 4 operators (with $p=0,1,2$ and 3) in the $0^{-+}$ and $1^{--}$ channels to get radially excited states. 

\begin{table}
\renewcommand{\arraystretch}{1.2}  
\caption{
  The meson/hybrid spin-singlet operators used in each channel.
 \label{table:mesonops}}
\begin{center}
\begin{tabular}{clcc}
 $J^{PC}$ & & Degeneracies & Operator \\ \hline
 $0^{-+}$ & ${\rm S}$& $1^{--}$ &$\tilde\chi^\dag\ \left[\tilde\Delta^{(2)}
     \right]^p\ \tilde\psi$ \\
 $1^{+-}$ & ${\rm P}$& $0,1,2^{++}$ &$\tilde\chi^\dag
    \ \tilde{\mbox{\boldmath{$\Delta$}}} \ \tilde\psi$ \\
 $1^{--}$ & ${\rm H_1}$& $0,1,2^{-+}$ &
    $\tilde\chi^\dag\ \tilde{\bf B}
       \left[\tilde\Delta^{(2)}\right]^p\ \tilde\psi$ \\
 $1^{++}$ & ${\rm H_2}$& $0,1,2^{+-}$ & $\tilde\chi^\dag\ 
      \tilde{\bf B}\!\times\!\tilde{\mbox{\boldmath{$\Delta$}}}
      \ \tilde\psi$ \\
 $0^{++}$ & ${\rm H_3}$& $1^{+-}$ & $\tilde\chi^\dag\ \tilde{\bf B}
     \!\cdot\!\tilde{\mbox{\boldmath{$\Delta$}}}\ \tilde\psi$ \\\hline
\end{tabular}
\figfudge
\end{center}
\end{table}

The bare quark mass is set by matching the ratio ${\rm R=M_{\rm kin}^S/(E_{1P}-E_{1S})}$, where ${\rm M_{\rm kin}^S}$ is the kinetic mass of the ${\rm 1S}$ state, to its observed value $21.01(6)$. Several low statistics runs using a range of quark masses were done in order to tune the quark mass.  From the results of these runs, we estimate that the uncertainty in tuning the quark mass is about $5\%$. The scale is set using the average of the 1S-1P and 1S-2S splittings. We find $r_0^{-1}=450(15)$ MeV.

The masses in the $1^{+-}$, $1^{++}$, and $0^{++}$ channels are extracted by fitting the single correlators to a single exponential for sufficiently large ${\rm t}$. In each of the $0^{-+}$ and $1^{--}$ channels, the variationally optimized correlator matrix is fitted to obtain the radially excited states. The results are shown in Fig.~\ref{fig:scaling}.

\section{Born-Oppenheimer Approximation}

In the Born-Oppenheimer approach, the hybrid meson is treated analogous to a diatomic molecule: the slow heavy quarks correspond to the nuclei and the fast gluon field corresponds to the electrons. First, one treats the quark $\mrm{Q}$ and antiquark $\overline{\mrm{Q}}$ as spatially-fixed color sources and determines the energy levels of the excited gluon field as a function of the $\overline{\mrm{Q}}\mrm{Q}$ separation $\mrm{r}$; each of these excited energy levels defines an adiabatic potential $V_{\overline{\mrm{Q}}\mrm{gQ}}(\mrm{r})$.
The quark motion is then restored by solving the Schr\"odinger equation in each of these potentials. 

\begin{figure}
\begin{center}
\leavevmode
\epsfxsize=2.5in \epsfbox[60 250 650 700]{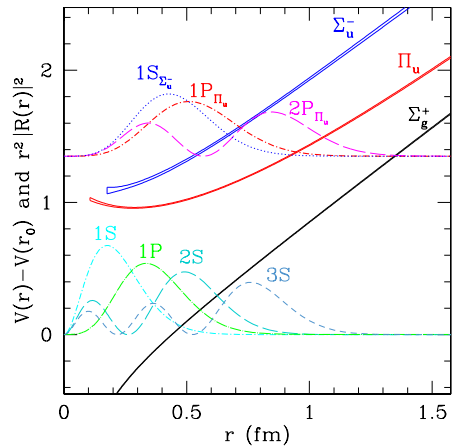}
\end{center}
\figfudge
\caption[]{
  Wavefunctions and potentials for the various hybrid/meson states.}
\label{fig:wavefunctions}
\end{figure}

The energy spectrum of the excited gluon field in the presence of a static quark-antiquark pair has been determined in previous studies\cite{earlier}. The three lowest-lying levels are shown in Fig.~\ref{fig:wavefunctions}. These levels correspond to energy eigenstates of the excited gluon field characterized by the magnitude $\Lambda$ of the projection of the total angular momentum ${\bf J}_{\mrm{g}}$ of the gluon field onto the molecular axis; by $\eta=\pm 1$, the symmetry quantum number under the combined operations of charge conjugation and spatial inversion about the midpoint between the quark and antiquark; and for $\Lambda=0$ states, a reflection in a plane containing the molecular axis. 

Given these static potentials, the LBO spectrum and wavefunctions are easily obtained by solving the radial Schr\"odinger equation with a centrifugal factor
${\rm \langle {\bf L}_{\overline{Q}Q}^2\rangle = L(L+1)-2\Lambda^2+\langle {\bf J}_g^2 \rangle}$ where ${\rm {\bf L}_{\overline{Q}Q}}$ is the orbital angular momentum of the quark--antiquark pair and ${{\rm \bf L}}={\rm {\bf L}_{\overline{Q}Q}}+{\rm {\bf J}}_g$. For the $\Sigma_{\mrm{g}}^+$ potential,
$\langle {\bf J}_{\mrm{g}}^2 \rangle=0$. The angular momentum of the excited glue is constrained by $\langle {\bf J}_{\mrm{g}}^2 \rangle\ge\Lambda(\Lambda+1)$. For the $\Pi_{\mrm{u}}$ levels, we attribute the lowest value and for the $\Sigma_{\mrm{u}}^-$ levels, we take $\langle {\bf J}_{\mrm{g}}^2 \rangle=2$ which is suggested by models of QCD. We show the effect of the uncertainty in this quantity by changing the value by one unit. The parity ${\rm P}$ and charge conjugation ${\rm C}$ of each meson is given in terms of ${\rm L}$ and ${\rm S}$ by ${\rm P = \epsilon\ (-1)^{L+\Lambda+1}}$ and ${\rm C = \epsilon\ \eta\ (-1)^{L+\Lambda+S}}$, where ${\rm L\geq \Lambda}$ and $\epsilon=1$ for $\Sigma^+$, $\epsilon=-1$ for $\Sigma^-$, and $\epsilon=\pm 1$ for $\Lambda>0$.

\begin{figure}
\begin{center}
\leavevmode
\epsfxsize=2.5in \epsfbox[60 200 600 700]{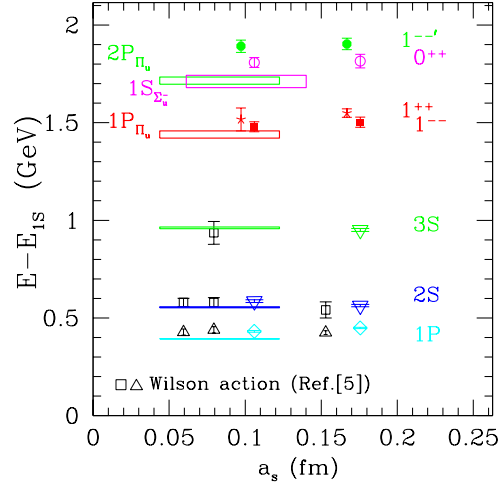}
\end{center}
\figfudge
\caption[]{
  Energy splittings with respect to the 1S state in GeV.
  The boxes are the Born-Oppenheimer spectrum where the height
  of the boxes show the effect of $\delta <J_g^2>=1$.}
\label{fig:scaling}
\end{figure}

Results for the LBO spectrum (splittings) of conventional ${\rm \overline bb}$ and hybrid ${\rm\overline{b}gb}$ states are shown in~Fig.~\ref{fig:scaling}. The scale $r_0$ is taken from the NRQCD simulation. The heavy quark mass ${\rm M_b}$ is tuned to reproduce the experimentally-known ${\rm \Upsilon(1S)}$ mass: ${\rm M_\Upsilon=2M_b+E_0}$, where ${\rm E_0}$ is the energy of the lowest-lying state in the $\Sigma_{\mrm{g}}^+$ potential. Level splittings are insensitive to small changes in the heavy quark mass. For example, a $5\%$ change in ${\rm M_b}$ results in changes to the splittings (with respect to the $\mrm 1S$ state) ranging from $0.1-0.8\%$.

\begin{table}
\caption[tabone]{
  Comparison between LBO and NRQCD splittings in GeV.
  The first error in LBO is due to a $5\%$ change in $m_b$ and the
  second error is an estimate of the uncertainty due to $\delta<J_g^2>$.
  \label{table:results}}
\begin{center}
\begin{tabular}{cllc}
                          & LBO            & NRQCD       & \% diff \\\hline
$\rm{1P}-\rm{1S}$         & $0.392(1)$     & $0.432(4)$  & $9(1)$ \\
$\rm{2S}-\rm{1S}$         & $0.553(5)$     & $0.586(5)$  & $6(1)$ \\
$\rm{3S}-\rm{1S}$         & $0.957(9)$     & $0.950(8)$  & $1(1)$ \\[1mm]
$\rm{H_1}-\rm{1S}$        & $1.421(8)(37)$ & $1.479(24)$ & $4(3)$ \\
$\rm{H_2}-\rm{1S}$        & $1.421(8)(37)$ & $1.517(58)$ & $6(5)$ \\
$\rm{H_3}-\rm{1S}$        & $1.711(5)(63)$ & $1.808(25)$ & $5(4)$ \\
$\rm{H_1^\prime}-\rm{1S}$ & $1.697(2)(36)$ & $1.892(30)$ & $10(2)$\\\hline
\end{tabular}
\figfudge
\end{center}
\end{table}

The applicability of the leading Born Oppenheimer approximation relies on the smallness of the retardation effects. We can determine the size of this effect by comparing the two approaches taken here since the NRQCD Hamiltonian only differs from the LBO Hamiltonian by the inclusion of the ${\rm \vec{p}\cdot\vec{A}}$ coupling between the color charge {\em in motion} and the gluon field. The difference in splittings are tabulated in Table~\ref{table:results} and shown in Fig.~\ref{fig:scaling}. In the comparison, the scale ambiguity has been removed and the lattice artifacts were found to be small ($<4\%$). Furthermore, the splittings are insensitive to small changes in the $b$-quark mass. We thus conclude that the small differences ($<\!10\%)$ from the two approaches is in fact due to retardation, hence validating the Born-Oppenheimer expansion. 

\section{Conclusion}

The masses of 4 hybrid states were determined in leading order NRQCD in the quenched approximation. The magnitude of the retardation effects were found to be small validating the Born-Oppenheimer expansion. 

The level splittings in the LBO approximation were found to be rather insensitive to the b quark mass. The LBO wavefunctions revealed the spatial largeness of the hybrid meson states and also indicated that the finite volume effects in our simulations should be negligible. Note that our LBO spectrum of hybrid mesons based on adiabatic surfaces calculated in the quenched approximation will very likely differ from the true spectrum due to our neglect of sea quark effects. The inclusion of such effects remains an important challenge for the future.


\begin{thebibliography}{9}
\bibitem{hasenfratz}
   P.~Hasenfratz {\it et al.},
   Phys.\ Lett.\ B {\bf 95}, 299 (1980);
   D.~Horn and J.~Mandula, Phys.\ Rev.\ D {\bf 17}, 898 (1978).
\bibitem{michael}
   S.~Perantonis and C.~Michael, Nucl.\ Phys.\ {\bf B 347}, 854 (1990).
\bibitem{peardon}
   C.~Morningstar and M.~Peardon, Phys.\ Rev.\ D {\bf 56}, 4043 (1997).
\bibitem{earlier}
   K.J.~Juge, J.~Kuti, and C.~Morningstar, Nucl.\ Phys. B
  (Proc.\ Suppl.) {\bf 63}, 326 (1998).
\bibitem{nrqcd}
   C.~Davies {\it et al.}, Phys.\ Rev.\ D {\bf 58}, 054505 (1998).
\end{thebibliography}
\end{document}